\documentclass[twocolumn,           % Format : preprint, twocolumn
               showpacs,            % Pacs : showpacs, noshowpacs
               preprintnumbers,     % Preprint: preprintnumbers,
               aps,                 % Society: ...
               prd,                 % Journal Style : pra, prb, prc, prd, pre,
               a4paper,             % Size : a4paper, ...
               groupedaddress,  % Affiliation (Title) : groupedaddress,
               nofootinbib,         % Footnote: footinbib, nofootinbib
               tightenlines,        % Remove additional spaces in a line
               floats,floatfix      % Floating pictures and tables
               ]{revtex4}

\usepackage{graphicx}
\usepackage{subfigure}
\usepackage{latexsym}
\usepackage{amsmath,amssymb}        % amssymb includes amsfonts
\usepackage[draft=false]{hyperref}
\usepackage{enumerate}
\usepackage{color}

\newcommand{\params}{\bar{\theta}}
\newcommand{\model}{\mathcal{M}}
\newcommand{\be}{\begin{equation}}
\newcommand{\ee}{\end{equation}}

\begin{document}

%%--- DRAFTCOPY --------------------------------
%% Prints a large "DRAFT" diagonally across each page
%% Does not show up in TeXview
%% \typeout{Prints "DRAFT" on each page; does not show in TeXView}
%% \special{!userdict begin /bop-hook{gsave 200 30 translate
%% 65 rotate /Times-Roman findfont 216 scalefont setfont
%% 0 0 moveto 0.90 setgray (DRAFT) show grestore}def end}
%%------------------------------------------------

%======================================%
%<<<<<<<<<<<< TITLE PAGE >>>>>>>>>>>>>>%
%======================================%

\title{Application of Bayesian model averaging to measurements of the
  primordial power spectrum}

\author{David Parkinson} 
\affiliation{Astronomy Centre, University of Sussex, Brighton BN1 9QH, 
United Kingdom}  
\author{Andrew R.~Liddle} 
\affiliation{Astronomy Centre, University of Sussex, Brighton BN1 9QH, 
United Kingdom} 
\date{\today} 
\pacs{98.80.-k}
%\preprint{arxiv} 
 
%======================================% 
%<<<<<<<<<<<<< ABSTRACT >>>>>>>>>>>>>>>% 
%======================================% 
 
\begin{abstract} 
Cosmological parameter uncertainties are often stated assuming a
particular model, neglecting the model uncertainty, even when Bayesian
model selection is unable to identify a conclusive best model.
Bayesian model averaging is a method for assessing parameter
uncertainties in situations where there is also uncertainty in the
underlying model.  We apply model averaging to the estimation of the
parameters associated with the primordial power spectra of curvature
and tensor perturbations. We use CosmoNest and MultiNest to compute
the model evidences and posteriors, using cosmic microwave data from
WMAP, ACBAR, BOOMERanG and CBI, plus large-scale structure data from
the SDSS DR7. We find that the model-averaged 95\% credible interval
for the spectral index using all of the data is $0.940 < n_{\rm s} <
1.000$, where $n_{\rm s}$ is specified at a pivot scale $0.015 \, {\rm
  Mpc}^{-1}$. For the tensors model averaging can tighten the credible
upper limit, depending on prior assumptions.
\end{abstract} 
 
\maketitle 
 
%======================================% 
%<<<<<<<<<<<<<< ARTICLE >>>>>>>>>>>>>>>% 
%======================================% 
 
\section{Introduction} 

Measurements of the cosmic microwave background (CMB) by the Wilkinson
Microwave Anisotropy Probe (WMAP) have, over the last few years
\cite{wmap3,wmap5,wmap7}, produced ever more refined constraints on
the cosmological parameters, particularly those relating to the
spectrum of primordial perturbations. Typically, when constraints are
quoted this is done under the assumption that a particular underlying
model is the correct one, but there remains some uncertainty as to
which is the most appropriate model to fit to the data. One approach
to this is model selection, which asks the data to rank the models
under consideration, but current applications
\cite{MPL,PML,RT2007,PPS_ev,Kawasaki:2009,Kilbinger:2009} indicate that
several models are still allowed.
 
Such uncertainty in the correct choice of model can be handled by the
technique of Bayesian model averaging \cite{hoeting}, which allows one
to assess parameter uncertainties in the presence of model
uncertainty. The individual posteriors from different models will
contribute to the model-averaged posterior, weighted by their model
likelihood.  Our aim in this paper is to apply this technique to
measurements of the primordial spectrum.  In Section \ref{sec:bayes}
we review the Bayesian ideas we are applying to measurements of the
primordial power spectrum. In Section \ref{sec:apps} we carry
out the analysis.

\section{Multi-model Bayesian statistics} 
\label{sec:bayes} 
 
Bayes' theorem states the relationship between models ($\model$),
parameters ($\params$) and data ($D$)  
\be
P(\params|D,\model) =
\frac{P(D|\params,\model)P(\params|\model)}{P(D|\model)} \,, 
\ee 
where $P(\params|\model)$ is the prior probability distribution of the
parameters (assuming some model), $P(D|\params,\model)$ is the
likelihood, and $P(\params|D,\model)$ is the posterior probability
distribution of the parameters. The prior is updated to the posterior
by the likelihood. The term $P(D|\model)$ represents the model
likelihood, and is called the {\it evidence}. In the case of
single-model inference (where only a single model or set of parameters
is considered), the evidence is simply a normalizing constant, set to
satisfy the condition that the posterior distribution sums to unity.

However, in most interesting cases in cosmology, the correct model is
not known, and the evidence takes different values for different
models. We can use Bayes' theorem again, at one level above, to
calculate the posterior odds between different models and perform {\it
  model selection}, 
\be
\frac{P(\model_1|D)}{P(\model_2|D)} =
\frac{P(D|\model_1)}{P(D|\model_2)} \frac{P(\model_1)}{P(\model_2)}
\,. 
\label{eq:modsel}
\ee
Here $\model_1$ and $\model_2$ are the different models under
consideration, $P(\model_i)$ gives the prior probability of each
model, and $P(\model_i|D)$ is the model posterior probability.  Thus
the model posterior probability is updated from the prior by the
evidence.  If the model priors are equal then the ratio of posteriors
is simply the ratio of evidences.  The ratio of evidences is commonly
known as the Bayes factor $B$ \cite{KassRafterty}, and an
interpretation scale was suggested by Jeffreys \cite{Jeff} (though
some authors have started to use different language to qualify the
different levels, e.g.\ Ref.~\cite{RT2007}). Many papers have already
been written about the use of the evidence for cosmological model
selection
\cite{ev,MPL,PML,RT2007,PPS_ev,Kawasaki:2009,Kilbinger:2009}.

The logical procedure would be to first perform a model selection
analysis to find the best model. Having done so, we would then perform
parameter inference for the parameters of that single best
model. However it is possible, even likely, that no model will have
decisive evidence ($\ln B > 5$) over all competing models. If we want
to include this model uncertainty in the parameter posteriors, we
could instead produce a {\it model-averaged} posterior distribution
\cite{hoeting}, where the individual posteriors from each model are
summed together, weighted by the model posterior values,
\be P(\params|D) = \frac{\sum_k
  P(\params|D,\model_k)P(\model_k|D) } {\sum_k P(\model_k|D)} \,.  
\ee
This model-averaged posterior encodes the uncertainty as to the
correct model.\footnote{Though it may be that in the end the `true'
  model is not even one that we have considered at the time of the
  analysis. ``Essentially, all models are wrong, but some are
  useful.'' \cite{box}} This model-averaging procedure has been used
before in cosmology \cite{BMA} and astrophysics/geophysics
\cite{BMAastro}.
 
\section{Application to Data} 
\label{sec:apps} 

The primordial power spectrum of scalar perturbations is normally
modeled through a modified power-law function of wavenumber $k$, 
\be
\Delta^2_{\mathcal{R}}(k) =
\Delta_{\mathcal{R}}^2(k_*)\left(\frac{k}{k_*}\right)^{(n_{\rm s}-1)+\frac12
  \ln(k/k_*)n_{\rm run}} \,, 
\ee
where the amplitude is defined at a pivot scale ($k_*$), $n_{\rm s}$
is the spectral index (also known as the tilt), and $n_{\rm run}$ is
the running of the spectral index. We also refer to
$\Delta_{\mathcal{R}}^2$ at the pivot scale as $A_{\rm s}$. A
maximally-symmetric Harrison--Zel'dovich (HZ) \cite{HZ} model has
equal power on all scales, so the spectral index is unity and the
running is zero. We discuss the choice of $k_*$ below.

Inflation, currently our best model for the generation of the spectrum
of Gaussian, adiabatic superhorizon perturbations, additionally predicts a
spectrum of tensor perturbations, which are also modeled through a
power law, 
\be
\Delta^2_h(k) = \Delta^2_h(k_*) \left(\frac{k}{k_*}\right)^{n_{\rm T}} \,,
\ee
where  $n_{\rm T}$ is the spectral index of the tensor perturbations (the
tensor running is normally neglected). Single-field, slow-roll
inflation predicts a consistency relation between the scalar and
tensor amplitudes (measured at the same scale) in terms of the tensor
spectral index,  
\be
\frac{\Delta_h^2(k_*)}{\Delta_{\mathcal{R}}^2(k_*)} \equiv r = -8n_{T} \,,
\ee
and we will enforce this throughout.

\begin{table}
\centering
\begin{tabular}{|l|l|c|c|c|}  \hline
Models & Parameter & Min & Max \\ \hline
All & $\Omega_{\rm b} h^2$ & 0.018 & 0.032 \\
 & $\Omega_{\rm c} h^2$ &  0.04 & 0.16 \\
& $\theta$ &  0.98 & 1.1 \\
& $\tau $ &  0.01 & 0.3 \\
& $\ln[10^{10} A_{\rm s}]$ & 2.6 & 4.2 \\
& $A_{\rm SZ}$ & 0 & 2 \\ \hline
tilt, tensor, run, tensor+run & $n_{\rm s}$ &  0.8 & 1.2 \\ \hline
tensor, tensor+run & $r$ &  0 & 1 \\ \hline
 run, tensor+run & $n_{\rm run}$ & -0.1 & 0.1 \\ \hline
\end{tabular}
\caption{\label{table:prior} Prior ranges for the parameters in the
  different models. We considered only uniform priors in this
  analysis. The priors on power spectrum parameters are set at $k_0=0.05\,
  {\rm Mpc}^{-1}$, the default for CosmoMC, but are so wide compared
  to the posteriors that the subsequent translation to the pivot scale
  is unaffected.}  
\end{table}

We considered five different models of the spectrum of primordial
perturbations in this analysis:  
\begin{enumerate}[I.]
\item A scale-invariant HZ spectrum of scalar
  perturbations  with no tensor component ($n_{\rm s}=1$, $r=0$). 
\item A tilted model, where the spectral index is allowed to vary,
  still with no tensors. 
\item A running model, where both the spectral index and the running
  of the spectrum ($n_{\rm run}$) are allowed to vary. 
\item A tensor model, where the spectral index of the scalar
  perturbations and the tensor-to-scalar amplitude ratio ($r$) are
  allowed to vary. 
\item A tensor+running model, where the spectral index,
  tensor-to-scalar ratio, and running all vary. 
\end{enumerate}
The priors on the parameters in these models are given in Table
\ref{table:prior}.

In this analysis we used measurements of the CMB temperature and
polarization power spectra from both the WMAP 5yr \cite{wmap5} and
7yr \cite{wmap7} releases, to explore how WMAP has improved in its
ability to distinguish between different models of the primordial
power spectrum. A compilation of WMAP 7yr and ground-based CMB
experiments (ACBAR \cite{ACBAR}, CBI \cite{CBI} and BOOMERanG
\cite{boom}), along with the Sloan Digital Sky Survey (SDSS) Data
Release 7 \cite{sdssdr7} measurements of the galaxy clustering power
spectrum, was also studied.

We used nested sampling \cite{Skilling} to compute the evidence values
and posterior distributions for the different models, making use of
CosmoNest \cite{MPL} and MultiNest \cite{Multinest} as additional
modules for the CosmoMC \cite{Lewis:2002} analysis code.

\begin{table}
\centering
\begin{tabular}{|c|c|c|c|c|}
\hline
\multicolumn{2}{|c|}{Model} & \multicolumn{3}{c|}{Datasets} \\
\multicolumn{2}{|c|}{} & WMAP 5yr & WMAP 7yr & WMAP 7yr+ext \\ \hline
I. & HZ  &    $ ~~0.0 \pm 0.1~$  &   $ ~~0.0 \pm 0.2~$ & $ 0.0 \pm 0.2$  \\
II. & $n_{\rm s}$ & $ ~~0.5 \pm 0.2~$ & $~~1.0 \pm 0.2~$ & $3.0 \pm 0.2$ \\
III. & $n_{\rm s}$+$n_{\rm run}$ & $ -0.1 \pm 0.2~$ & $~~0.4 \pm 0.2~$ & $3.4 \pm 0.2$ \\
IV. & $n_{\rm s}$+$r$ & $ -1.3 \pm 0.1~$ & $-0.9 \pm 0.2~$ & $0.8 \pm 0.2 $ \\
V. & $n_{\rm s}$+$n_{\rm run}$+$r$ &  $-1.1 \pm 0.2~$     & $-0.7 \pm 0.2~$ &
$2.2 \pm 0.2$ \\ \hline 
\end{tabular}
\caption{\label{table:ev_diff} (log-)evidence differences for the
  different models, compared to the HZ model for that data
  compilation. Positive values mean that the model is favored over
  HZ. The unnormalized evidence values for the HZ model are -1346.3
  (for WMAP 5yr), -3754.5 (for WMAP 7yr) and -3834.3 (for WMAP 7yr+ext).} 
\end{table}

\begin{figure*}[ht!]
\includegraphics[width=0.8\linewidth]{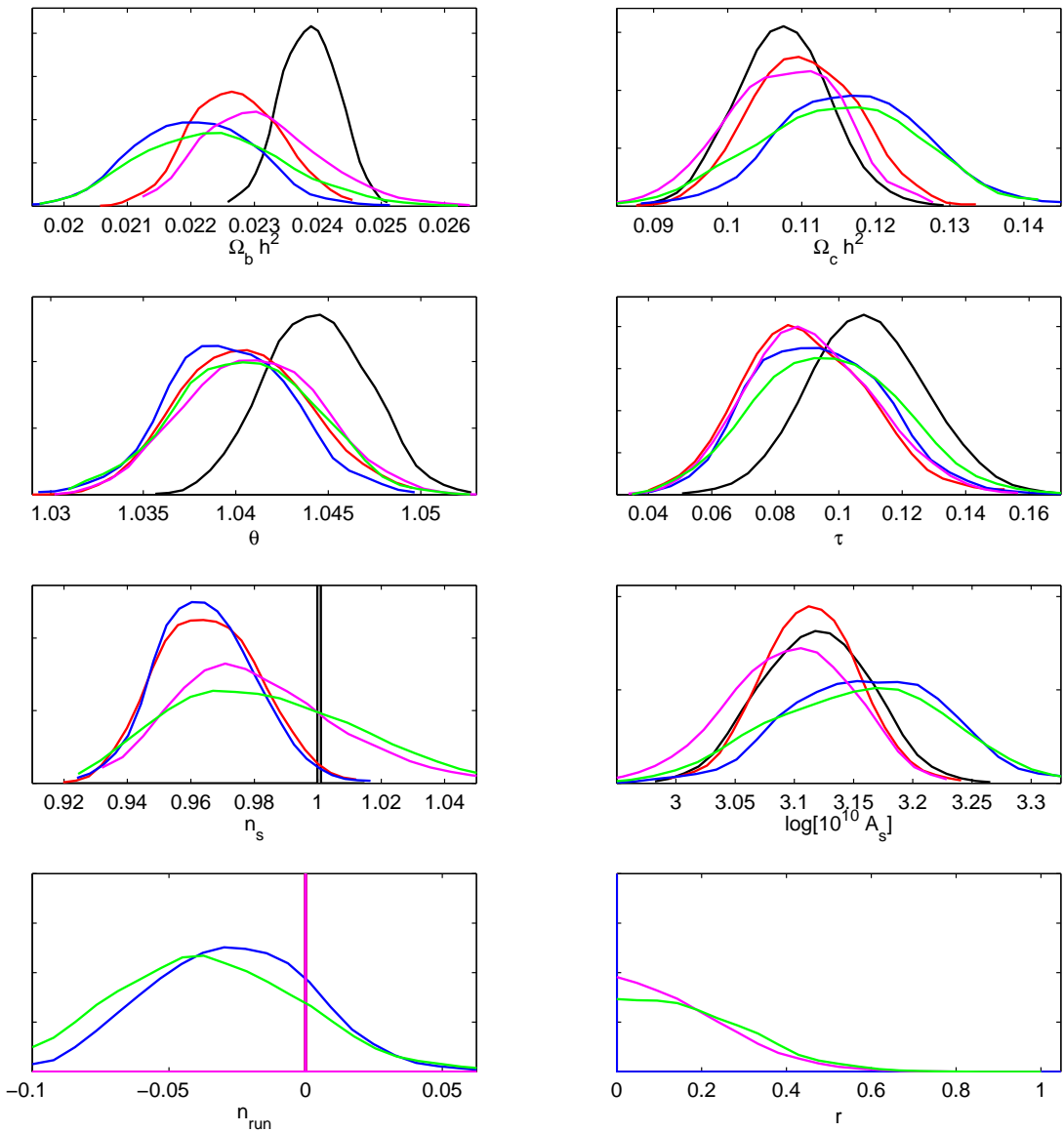}
\caption{The posterior probability distributions for the different
  models, using only the WMAP 5yr data. The models are HZ (black),
  tilted (red), running $n_{\rm s}$+$n_{\rm run}$ (blue), 
  `inflation' $n_{\rm s}$+$r$ (magenta) and tensor+running (green).}   
 \label{fig:post_5yr}
 \end{figure*}
 
The evidence values for the various models and different data
compilations are given in Table \ref{table:ev_diff}. We find that the
tilted model is the favored model for all data compilations, except
for WMAP 7yr+ext where it is tied with the tensor+running model
within the evidence uncertainties. The tilted model is favored over
the HZ with strong, but not decisive, evidence (as defined in the
Jeffrey's scale) using the combined data sets. The tensor model is
disfavored compared to the HZ using WMAP 5yr, and 7yr data alone, but
becomes mildly favored when the other data is included (though the
evidence differences are too small to be conclusive). The running
model has approximately the same evidence as the HZ for WMAP 5yr and
7yr, but this increases to be about the same as the tilted model when
the extra datasets are added.

\begin{figure*}[t]
\includegraphics[width=0.8\linewidth]{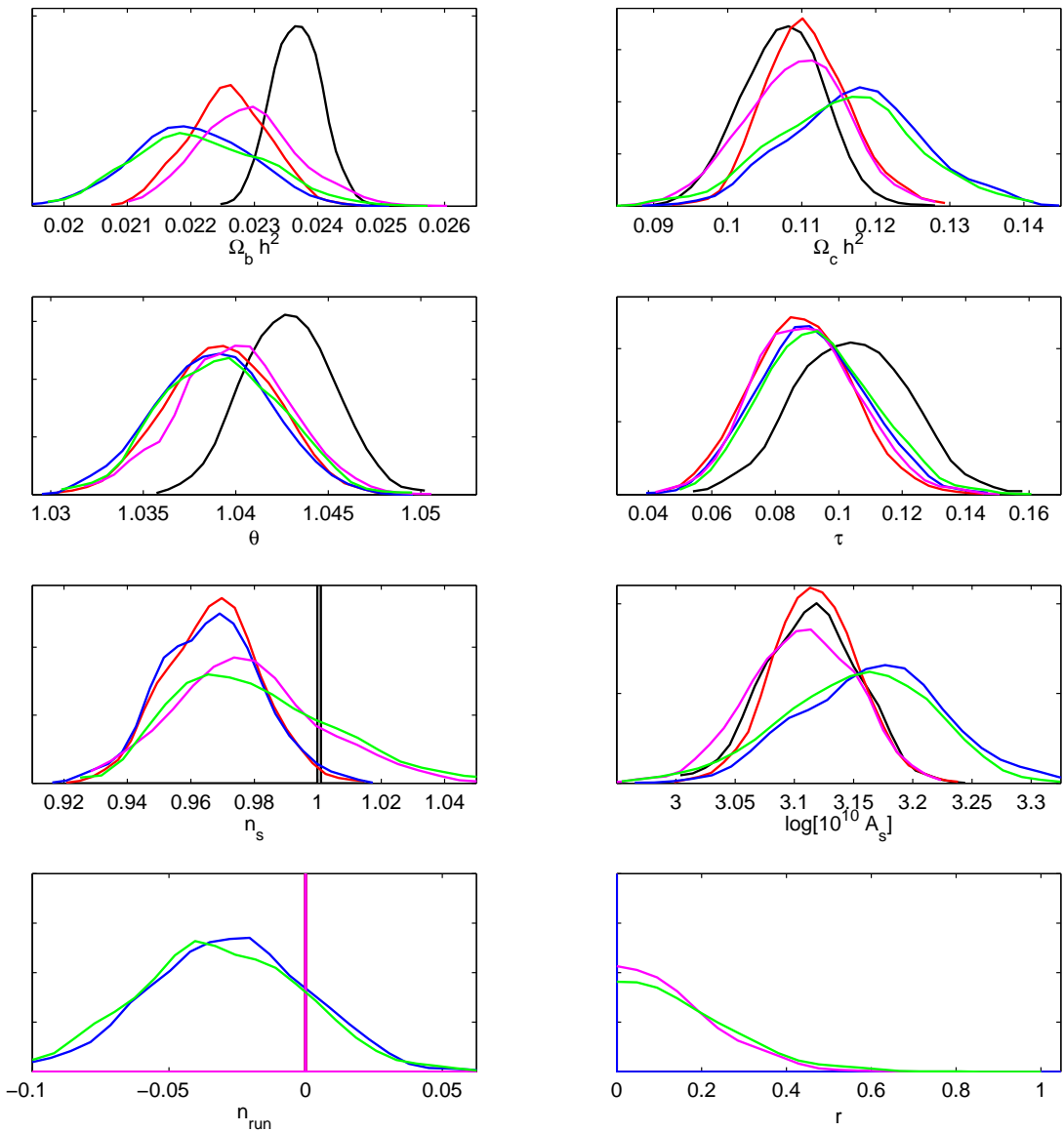}
\caption{The posterior probability distributions for the different
  models, using only the WMAP 7yr data. The models are HZ (black),
  tilted (red), running $n_{\rm s}$+$n_{\rm run}$ (blue),
  `inflation' $n_{\rm s}$+$r$ (magenta) and tensor+running (green).}   
 \label{fig:post_7yr}
 \end{figure*}

The question of what the probability of $n_{\rm s}=1$ (i.e.\ the probability
of the HZ model) is one of model selection. Combining the evidence
values with model prior probabilities, we can compute the normalized
model posterior for each of the models, using a normalized version of
Eq.~(\ref{eq:modsel})
\begin{equation}
P(\model_i|D) = \frac{P(D|\model_i)P(\model_i)}{\sum_k
  P(D|\model_k)P(\model_k)}\,. 
\end{equation}
Assuming equal prior probabilities for each of the models, $P({\rm
  HZ})=0.24$ for WMAP 5yr, $0.164$ for WMAP 7yr and $0.016$ for WMAP
7yr+ext. So even WMAP 7yr data is not strong enough to exclude an HZ
spectrum by itself, but the addition of extra data reduces its model
probability by a factor of 10, reducing it to less than 2\%.

Other recent papers \cite{Kawasaki:2009,Kilbinger:2009} have also
computed the Bayes factors for models of the primordial power
spectrum, using the WMAP 5yr data and other extra data sets. Though it
is difficult to do a direct comparison of raw evidence values between
our analysis and others (owing to slightly different choices of
priors), their basic conclusions are the same as ours: tilted models
are favored over HZ, a tensor model is disfavored relative to the
others, and running models have roughly the same evidence as tilted
models when other data is added in.

%\begin{table*}
%\centering
%\begin{tabular}{|c|c|c|c|}
%\hline
%Model & \multicolumn{3}{c|}{Datasets} \\
% & WMAP-5 & WMAP-7 & WMAP-7+ext \\ \hline
%HZ  &    $ -1346.31 \pm 0.10$  &   $-3754.54 \pm 0.20$ & $-3834.33
%\pm 0.21$  \\ 
%$n_s$ & $-1345.85 \pm 0.16$ & $-3753.57 \pm 0.21$ & $-3831.38 \pm 0.22$ \\
%$n_s$+$r$ & $-1347.57\pm 0.11$ & $-3755.41 \pm 0.22$ & $-3833.54
%\pm 0.23 $ \\ 
%$n_s$+$n_{\rm run}$ & $ -1346.41 \pm 0.15$ & $-3754.19 \pm 0.22$ &
%$-3830.96 \pm 0.22$ \\ 
%$n_s$+$n_{\rm run}$+$r$ &  $-1347.38 \pm 0.18$     & $-3755.22 \pm
%0.22$ & $-3832.15 \pm 0.23$ \\ \hline 
%\end{tabular}
%\caption{\label{table:ev_vals} (log-)evidence values for the
%different models and  
%different data sets, as discussed in the text.}
%\end{table*}

\begin{figure*}[t]
\includegraphics[width=0.8\linewidth]{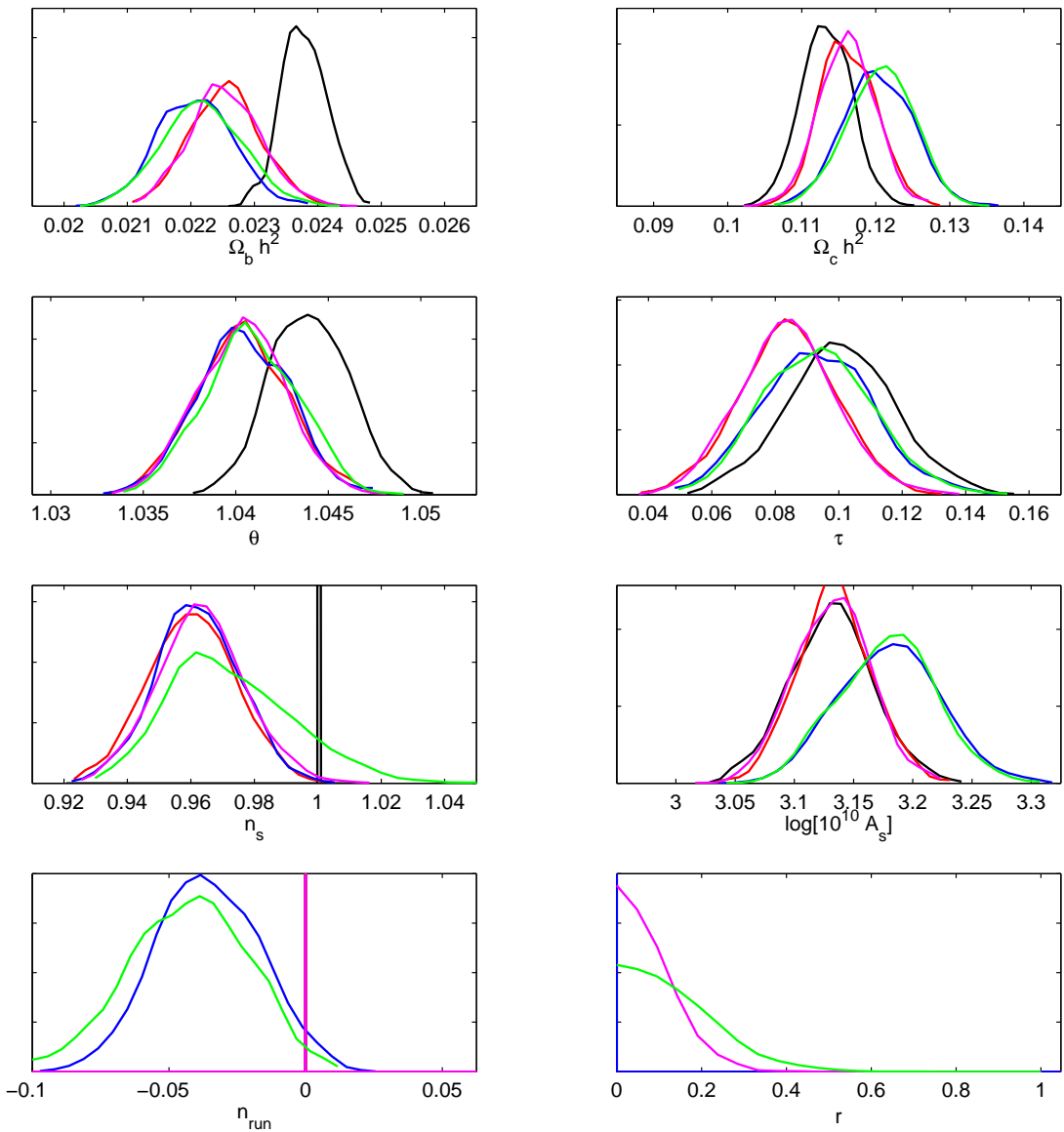}
\caption{The posterior probability distributions for the different
  models, using only the WMAP 7yr data plus other datasets. The
  models are HZ (black), tilted (red),  running 
  $n_{\rm s}$+$n_{\rm run}$ (blue), `inflation' $n_{\rm s}$+$r$
  (magenta) and tensor+running (green).} 
 \label{fig:post_7yr+ext}
 \end{figure*}
 
The one-dimensional marginalized posteriors for each parameter, model
and data compilation are shown in Figs.~\ref{fig:post_5yr} (for WMAP
5yr), \ref{fig:post_7yr} (for WMAP 7yr), and \ref{fig:post_7yr+ext}
(for WMAP 7yr+ext). The posteriors are normalized in the usual way,
such that the area under each curve is unity. For models where a
parameter is not varied (such as the spectral index in the HZ model),
a delta function at the appropriate parameter value is the relevant
posterior.

In terms of plotting constraints on the power spectrum parameters, the
choice of pivot scale $k_*$ is important. If it is not optimized,
marginalized constraints can appear much weaker than they actually
are.  We chose the scale that decorrelates the uncertainty on the tilt
and the running (in the running model), following the method described
in Ref.~\cite{Cortes:2007ak}. This scale is found to be 0.013
Mpc$^{-1}$ for the WMAP 5yr and 7yr data and 0.015 Mpc$^{-1}$ for the
WMAP+ext dataset.

For the Bayesian model averaging, we assume that the prior model
probabilities are equal. Variation of this assumption could readily be
explored using the quoted evidence values; for instance one might want
to downweight HZ as it is not based on a physical model, or models
with running as inflationary models with large running are hard to
construct. Model averages are carried out by combining the posterior
samples from different models weighted by the appropriate model
probability. 

We do not show a model-averaged result for $A_{\rm s}$, as we did not
optimize the pivot scale for the amplitude; in the figures one sees
that the central amplitude is shifted in the running models. This
means that the amplitude is best determined at some other scale, and a
model averaging should only be carried out at that pivot scale if
constraining power is not to be lost. In any case one can see by eye
that model averaging will have little effect on the constraints on
$A_{\rm s}$, which is well determined in all models.

\begin{figure}[t]
\includegraphics[width=0.8\linewidth]{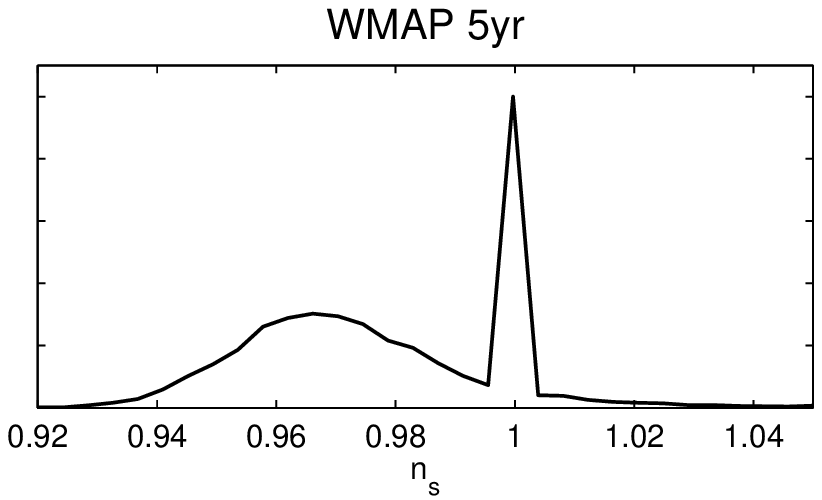}\\
\includegraphics[width=0.8\linewidth]{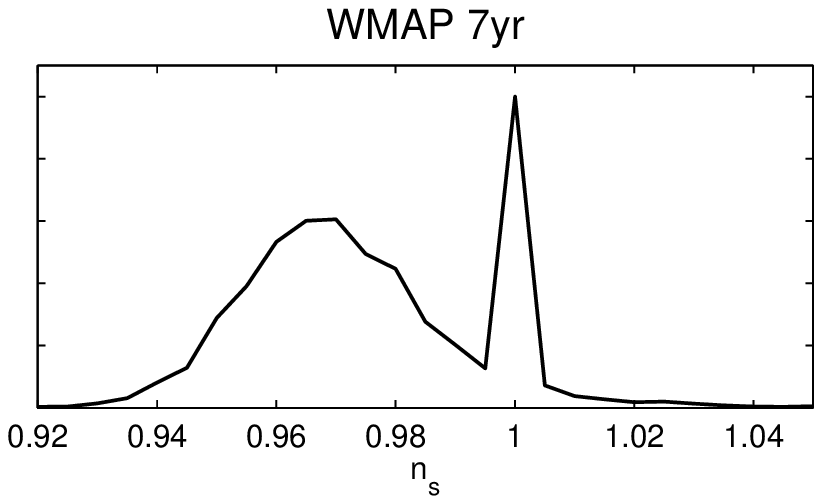}\\
\includegraphics[width=0.8\linewidth]{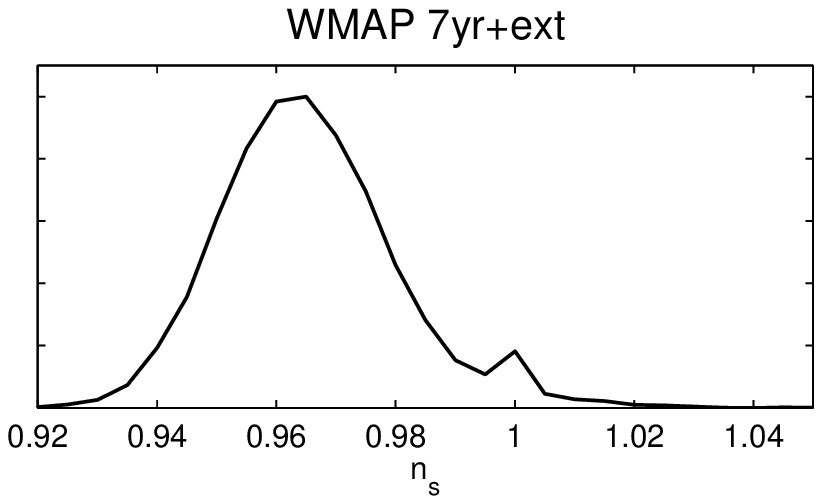}
\caption{The model-averaged posterior distributions for the spectral
  index $n_{\rm s}$, using the WMAP 5yr data only (top), the WMAP 7yr
  data only (middle) and the WMAP 7yr+ext compilation (bottom). The
  probability distribution includes a delta function around $n_{\rm
    s}=1$, artificially broadened in the plot by the
  binning process. }
\label{fig:averaged_5yr}
\end{figure}

The tilt $n_{\rm s}$ is more interesting, as it is not so well
determined due the residual probability that HZ is correct. Its
model-averaged posterior is shown in Fig.~\ref{fig:averaged_5yr}.
Note that in analyses with WMAP data alone the HZ `spike' is prominent
in the model-averaged posteriors, containing a significant fraction of
the posterior probability. Only once other data are brought in does
its effect become small. From the complete data compilation, we find
that the model-averaged limits on the tilt at the pivot scale are
$0.940 < n_{\rm s} < 1.000$ (95\% credible interval), the upper limit
being precisely at one as it happens to fall within the delta-function
component from the HZ model.

Variation of prior assumptions can modify these results somewhat.
Changes to the prior ranges of parameters common to all models, such
as $h$ and $\tau$, will have no effect, at least as long as the data
constrains the values to lie well within the prior as it does in these
cases. Modifying the priors on the parameters that are varied only in
some models can shift the results. As an example, we consider doubling
the prior range of $n_{\rm s}$, to $[0.6,1.4]$. As the added range
fits the data poorly, it has negligible likelihood and this halves the
evidence of models in which $n_{\rm s}$ varies, i.e.\ their log
evidences in Table~\ref{table:ev_diff} are reduced by $\ln 2 \sim
0.69$, which changes the quantitative outcome but not the qualitative
one. If one recomputes the 95\% confidence range of $n_{\rm s}$ under
this assumption the range is unchanged. Changes to the assumed prior
model probabilities can be handled similarly.

\begin{figure}[t]
\includegraphics[width=0.85\linewidth]{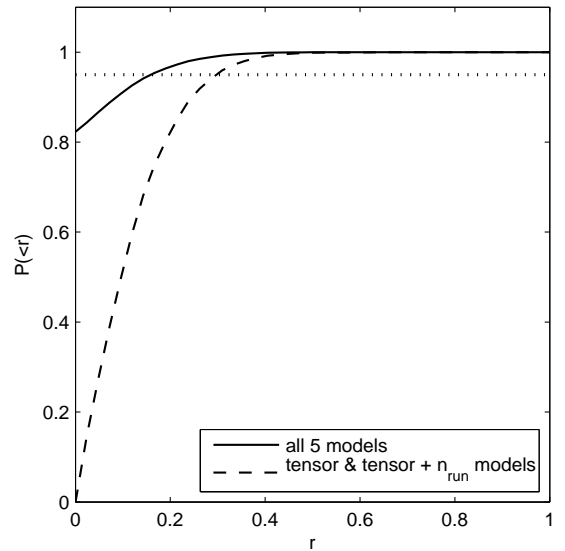} 
\caption{\label{fig:averaged_r} The model-averaged cumulative
  probability distribution for the tensor-to-scalar ratio $r$ using
  the WMAP 7yr+ext compilation. The solid curve gives the probability
  averaged over all five models, whereas the dashed curve gives the
  probability averaged just over those models where $r$ is varied
  (tensor and tensor+running). The dotted line shows the 95\% credible
  limit.}
\end{figure}

Finally, we consider the tensors. As they are not detected, model
uncertainty can have a significant impact by lending support to models
in which they are entirely absent, i.e.\ $r$ is precisely zero. If we
average over all five models, the model-averaged 95\% upper limit on
the tensor-to-scalar ratio is $r < 0.16$ (again at the pivot
scale). This is indeed somewhat tighter than results from individual
models (e.g.\ the equivalent upper limit from the `inflation' model is
$0.18$) because it allows for the possibility of no
tensors. Nevertheless, this result is clearly highly prior dependent,
and would for instance change if one decided that a logarithmic prior
on $r$ were more appropriate.\footnote{In practice a logarithmic prior
  on $r$ puts almost all the prior model probability at very small $r$
  values, yielding results near identical to a tensorless model provided
  tensors are not detected.}

An alternative tensor limit can be obtained by averaging only over the
two models which permit tensors, which gives $r<0.30$. The cumulative
model-averaged probabilities for $r$ under both assumptions are shown
in Fig.~\ref{fig:averaged_r}. It is clear that any upper limit quoted
on the tensor fraction has significant model and prior uncertainty, as
well as observational uncertainty.

\section{Conclusions}

We have illustrated the methodology of Bayesian model averaging using
primordial power spectrum estimation. From a Bayesian viewpoint, the
purpose of all data analysis is to start with prior information,
perhaps entirely theoretically motivated, and take data of increasing
quality until the data likelihood is able to convincingly overcome the
prior uncertainty. Bayesian model averaging allows us to include the
prior model uncertainty as well as the prior parameter uncertainty in
this process, and hence offers a more complete incorporation of
theoretical uncertainties.

Presently data are unable to decisively distinguish amongst different
models for the primordial perturbations, with all five that we discuss
remaining viable at some level. Despite that, parameters such as the
spectral amplitude that are very accurately measured are quite
unaffected by model uncertainty. Parameters moderately well
determined, such as $n_{\rm s}$, can see significant effects from
model uncertainty, while undetermined parameters such as $r$ are
naturally the most sensitive to the increased incorporation of
uncertainties.

Bayesian Model Averaging can be used beyond cosmological models to any
problem in cosmology or astrophysics where the underlying model is
uncertain. This may be uncertainty as to the nature of the physical
object (e.g. unresolved galaxies of different species contributing to
a background) or with regards to the data analysis (e.g. supernova
light curve analysis, where a number of different light curve fitters
are available).

As with any Bayesian analysis, the results will have some dependence
on the choices of priors, including the model prior probabilities. We
should not be afraid of this; the opportunity to choose appropriate
priors is our chance to deploy our physical intuition. Readers who
prefer different priors are welcome to recalculate if they wish; this
is particularly easy for model priors as the evidence ratios we quoted
are all that is needed. We also briefly discussed a modification of
parameter priors; a full analysis should explore the reasonable range
of prior possibilities.  Only by combining the full range of prior
uncertainties, at both parameter and model level, with observational
uncertainties can one obtain the full picture of current
understanding.

%======================================%
%<<<<<<<<<<< ACKNOWLEDGMENTS >>>>>>>>>>%
%======================================%

\begin{acknowledgments}

The authors were supported by the Science and Technology Facilities
Council [grant number ST/F002858/1]. We thank Mike Hobson, Martin
Kunz, and Roberto Trotta for helpful discussions. We acknowledge use
of the Archimedes computing cluster at the University of Sussex,
supported by funds from SRIF3, and the COSMOS supercomputer in
Cambridge, supported by SGI, Intel, HEFCE and STFC.

\end{acknowledgments}

%======================================% 
%<<<<<<<<<<<< BIBLIOGRAPHY >>>>>>>>>>>>% 
%======================================% 

%%%%%%%%%%%%%%%%%%%%%%%%%%%%%%%%%%%%%%%%%%%%%%%%%%%%%%%%%%%%%%%%%%%%%%%% 
\end{document}